\documentclass[sigconf]{acmart}
\usepackage{url}
\usepackage{textcomp}
\usepackage{times}
\usepackage{epsfig}
\usepackage[normalem]{ulem}
\usepackage{amsmath,amssymb,amsfonts}
\usepackage{xcolor}
\usepackage{graphicx}
\usepackage{listings}
\usepackage{color}
\usepackage{courier}
\usepackage{float}
\usepackage{algorithm} 
\usepackage[noend]{algpseudocode}
\usepackage[tight,footnotesize]{subfigure}
\usepackage{makecell}
\usepackage[capitalize]{cleveref}
\acmYear{2019}
\begin{document}
\title{Escoin: Efficient Sparse Convolutional Neural Network Inference on GPUs}
\author{Xuhao Chen}
\affiliation{University of Texas at Austin}
\email{cxh@utexas.edu}
\begin{abstract}
Deep neural networks have achieved remarkable accuracy in 
many artificial intelligence applications, e.g. computer vision,
at the cost of a large number of parameters and high computational 
complexity. Weight pruning can compress DNN models by removing 
redundant parameters in the networks, but it brings sparsity in the
weight matrix, and therefore makes the computation inefficient on GPUs.
Although pruning can remove more than 80\% of the weights, it actually 
hurts inference performance (speed) when running models on GPUs.

Two major problems cause this unsatisfactory performance on GPUs.
First, lowering convolution onto matrix multiplication reduces data
reuse opportunities and wastes memory bandwidth. Second, the 
sparsity brought by pruning makes the computation irregular, which
leads to inefficiency when running on massively parallel GPUs.
To overcome these two limitations, we propose Escort, an efficient
sparse convolutional neural networks on GPUs. Instead of using the
lowering method, we choose to compute the sparse convolutions directly.
We then orchestrate the parallelism and locality for the direct sparse
convolution kernel, and apply customized optimization techniques
to further improve performance. Evaluation on NVIDIA GPUs show that
Escort can improve sparse convolution speed by 2.63$\times$ and 3.07$\times$, 
and inference speed by 1.38$\times$ and 1.60$\times$, compared to 
CUBLAS and CUSPARSE respectively.
\end{abstract}


\maketitle
\section{Introduction}
Deep neural networks (DNNs)~\cite{DL} have been widely used in many artificial 
intelligence (AI) applications including computer vision~\cite{AlexNet,Video,Action}, 
speech recognition~\cite{Speech,DeepSpeech2}, natural language processing~\cite{NLP}, 
and robotics~\cite{DeepDriving,AlphaGo}. Modern DNNs are composed of five to 
more than a thousand network layers, with a trend of going deeper and more complex. 
A common form of DNNs is \textit{convolutional neural networks} (CNNs), which 
are mainly composed of multiple convolutional (CONV) layers. In recent CNNs, 
the CONV layers dominate the entire networks and consumes most of the execution 
time. This paper focuses on improving the speed of CONV layers in CNNs.

In many application domains, DNNs are now able to exceed human accuracy
~\cite{Tutorial}. The superior accuracy of DNNs, however, comes at the cost 
of high computational complexity. With continuous increase of their model sizes, 
DNNs consume considerable storage, memory bandwidth, and computational resources.
To address this limitation, weight pruning~\cite{prune} has been proposed 
to compress DNN models by removing redundant connections in the networks.
However, although this technique can significantly reduce the model size by 
removing an average of 80\% of the weights, pruning actually hurts inference 
performance (i.e. speed) when running CNN models on GPUs~\cite{Scalpel}. 

To understand the performance effect of weight pruning, we measured 
the inference speed of 3 popular CNNs on NVIDIA GPUs using 
CUBLAS~\cite{CUBLAS} and CUSPARSE~\cite{CUSPARSE} library respectively. 
In spite that weight pruning can remove a large portion of 
multiply-accumulate (MAC) operations, we discover that the inference 
speed of the networks using CUSPARSE is actually barely faster than that 
using CUBLAS. Two issues result in this performance degradation. 
First, the overhead of lowering convolution onto matrix multiplication 
becomes a severe problem when the computation turns into sparse after pruning.
The lowering approach has demonstrated overhead for dense convolution~\cite{CaffeconTroll},
since it duplicates the input features multiple times, which wastes memory bandwidth
and reduces the data reuse opportunities. 
For the dense case, this overhead is trivial. However, it becomes unacceptable 
for sparse convolution whose computational intensity is already much lower than 
dense convolution. For a highly memory bound operation like sparse convolution, 
lowering is no longer a suitable choice for implementing convolution on GPUs.

Second, sparse matrix computation is much less efficient than its 
dense counter part on GPUs. Although sparse matrix multiplication 
avoids unnecessary MAC operations, its memory access pattern is 
fairly irregular and can not fully take advantage of the compute 
capability of the GPU architecture. Besides, although sparse matrix 
computation using compressed data structure could save memory space, 
there is overhead to decode the sparse format at runtime.

To overcome the limitations, we propose Escort, an efficient sparse CNN 
method customized for GPU's data-parallel architecture. Instead of lowering 
the convolution onto matrix multiplication, we choose to directly compute 
the sparse convolution. To take advantage of GPU's tremendous computational 
horsepower, we customize the dataflow and apply a series of optimization 
techniques based on the understanding of the memory access pattern.
We implement Escort using CUDA and evaluate it on NVIDIA GPUs. Experimental
results show that Escort substantially outperforms the lowering method using 
either CUBLAS or CUSPARSE. To the best of our knowledge, this is the first 
direct sparse convolution tailored for the GPU architecture.
 
This paper makes the following contributions:

\begin{itemize}
\item We propose Escort, a direct sparse convolution approach that can efficiently
run on modern GPUs.

\item We orchestrate the parallelism and locality for Escort and optimize it for the 
GPU architecture.

\item We measure the inference speed of Escort on NVIDIA GPUs, and demonstrate 
its superior performance over CUBLAS and CUSPARSE.
\end{itemize}

The rest of the paper is organized as follows:
Section~\ref{sect:back} introduces the background of sparse convolutional 
neural networks and explains the motivation of this work.
Our proposed design is described in Section~\ref{sect:design}.
We present the evaluation in Section~\ref{sect:eval}.
Section~\ref{sect:relate} summarizes related works and
Section~\ref{sect:concl} concludes.

\begin{figure}[t]
	\begin{center}
		\includegraphics[width=0.48\textwidth]{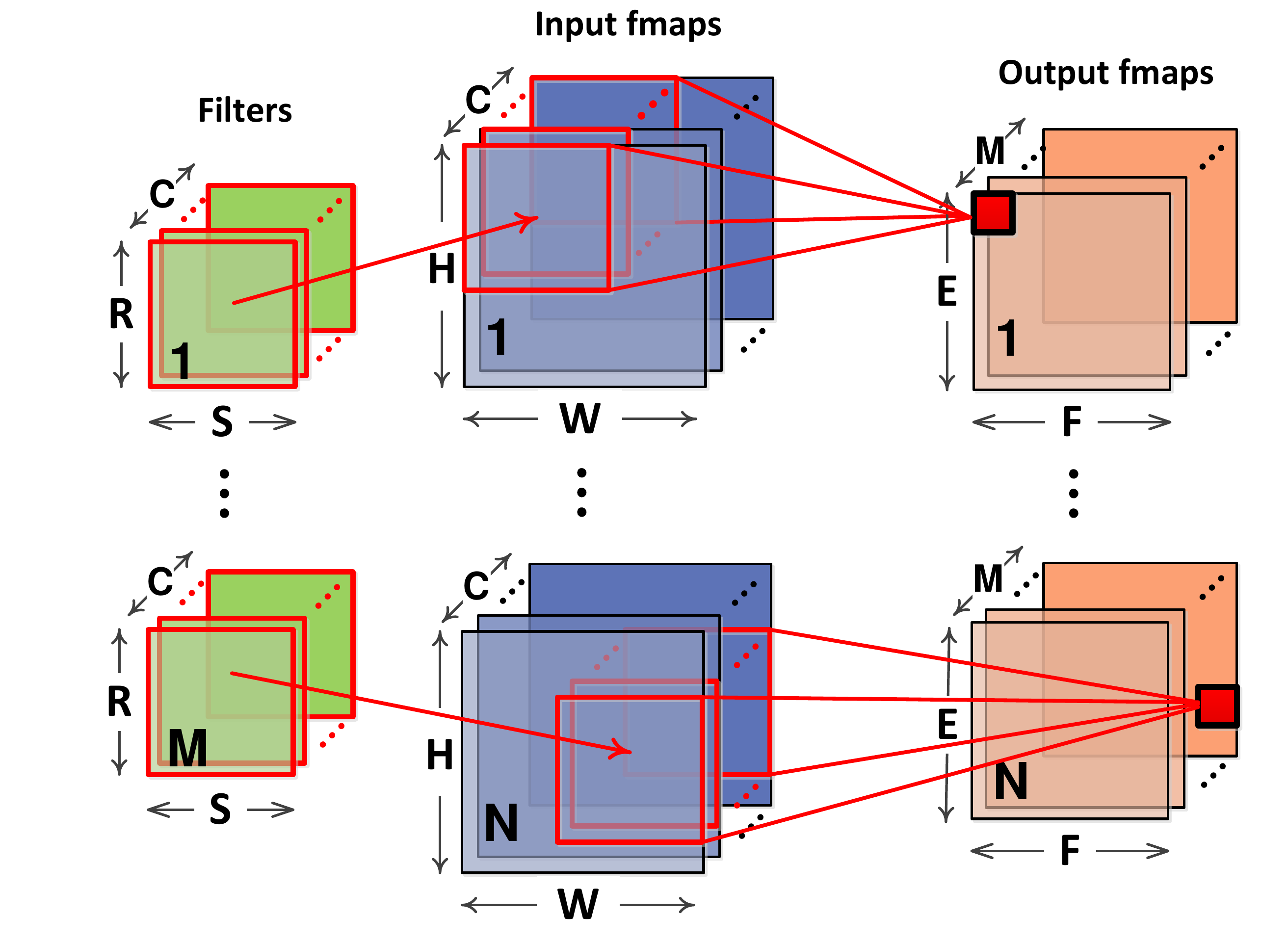}
		\vspace{-0.3cm}
		\caption{3-D convolutions in CNNs~\cite{Tutorial}.}
		\vspace{-0.8cm}
		\label{fig:cnn}
	\end{center}
\end{figure}

\section{Background and Motivation}\label{sect:back}

In AI applications, employing DNNs can be decomposed into two tasks: 
training and inference. Today, training is often done on GPUs, while 
inference depends on the application and can employ CPUs, GPUs, FPGAs 
or ASICs~\cite{Tutorial}. This paper focuses on CNN inference on GPUs. 
Since over 90\% of the computation of recent CNN designs is in 
convolutions~\cite{SparseCNN}, we tend to speed up this core operation.

\begin{algorithm}[!t]
\caption{Sequential Convolution~\cite{SCNN}}
\label{alg:conv}
\begin{algorithmic}[1]
\Procedure{Conv}{in, weight, out}
	\For{$n$ in [0, N)}
	\For{$m$ in [0, M)}
	\For{$c$ in [0, C)}
	\For{$h$ in [0, E)}
	\For{$w$ in [0, F)}
	\For{$r$ in [0, R)}
	\For{$s$ in [0, S)}
		\State out[$n$][$m$][$h$][$w$] += \\
			\hspace{4.0cm} in[$n$][$c$][$h$+$r$][$w$+$s$] * \\
			\hspace{4.0cm} weight[$m$][$c$][$r$][$s$]
	\EndFor
	\EndFor
	\EndFor
	\EndFor
	\EndFor
	\EndFor
	\EndFor
\EndProcedure
\end{algorithmic}
\end{algorithm}

\subsection{Convolutional Neural Networks}
Convolutional neural networks (CNNs)~\cite{CNN} have become the most 
popular algorithmic approach for deep learning in many application 
domains. Each of the CONV layers in a CNN is primarily composed of 
high-dimensional convolutions as visualized in \cref{fig:cnn}. 
In this computation, the core operation is a 2-D sliding window
convolution of an $R \times S$ filter kernel over a $H \times W$ 
input channel to produce a $E \times F$ output channel. 
A input feature map (ifmap) can include multiple ($C$) input channels. 
A distinct filter kernel is applied to each input channel, and 
the outputs for each of the $C$ channels are accumulated together 
element-wise into a single channel of output feature map (ofmap). 
Multiple 3-D filters ($M$) can be applied to the same 
volume of input activations to create $M$ output channels. 
Finally, $N$ ifmaps may be processed together as a batch to 
potentially improve reuse of the filter weights~\cite{Tutorial}. 

\begin{table}[b]
	\small
	\centering
	\begin{tabular}{c c}
		\hline
		\hline
		\bf{Shape Parameter} & \bf{Description}\\
		\hline
		N & {batch size}\\
		\hline
		M & {\# of filters / \# of ofmap channels}\\
		\hline
		C & {\# of ifmap/filter channels}\\
		\hline
		H/W & {ifmap height/width}\\
		\hline
		R/S & {filter height/width}\\
		\hline
		E/F & {ofmap height/width}\\
		\hline
		\hline
	\end{tabular}
	\caption{Shape Parameters of a CONV Layer~\cite{Tutorial}}
	\label{table:shape}
\end{table}

Given the shape parameters in Table~\ref{table:shape}, the computation 
of a CONV layer is defined as \cref{eq:cnn}, where $O$, $I$ and $W$ 
are the matrices of the ofmaps, ifmaps and filters, respectively. 
Filters are composed of weights, while input and output feature maps (ifmaps, 
ofmaps) are composed of activations. Algorithm~\ref{alg:conv} shows the 
pseudo code of computing a complete CONV layer. It is performed as a loop 
nest over 7 variables. Each point in the 7-dimensional space formed from these 
variables represents a single multiply-accumulate operation (line 9$\sim$11).

\begin{equation}
\label{eq:cnn}
\scriptsize
\begin{split}
&O[n][m][h][w] = \sum\limits_{c=0}^{C-1} \sum\limits_{r=0}^{R-1} \sum\limits_{s=0}^{S-1} I[n][c][h+r][w+s] \times W[m][c][r][s],\\
&0 \le n < N, 0 \le m < M, 0 \le h < E, 0 \le w < F,\\
&E = H - R + 1, F = W - S + 1.
\end{split}
\end{equation}

\begin{figure}[t]
	\begin{center}
		\includegraphics[width=0.36\textwidth]{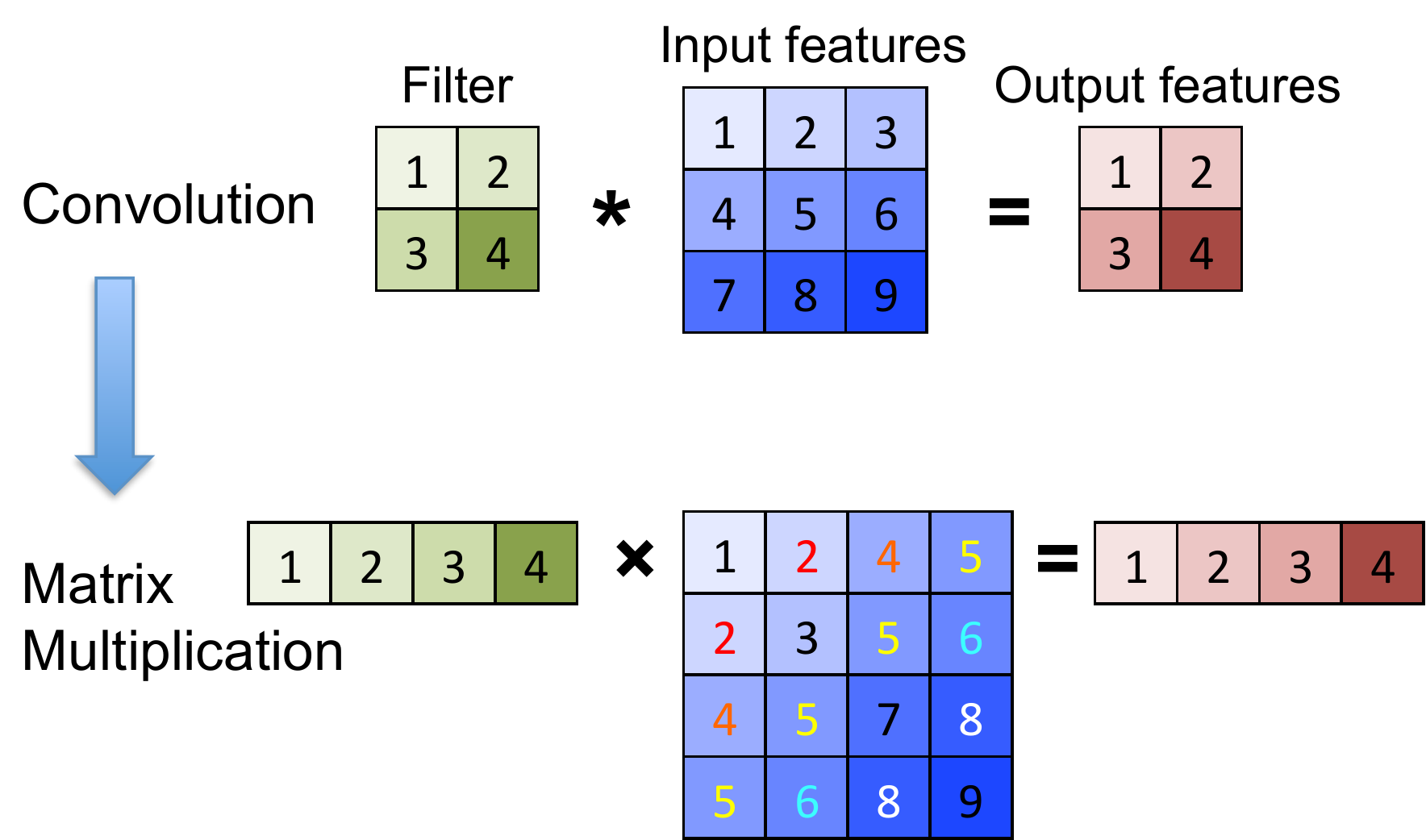}
		\caption{Lowering 2-D convolution to matrix multiplication~\cite{Tutorial}.}
		\vspace{-0.6cm}
		\label{fig:lowering}
	\end{center}
\end{figure}

\subsection{The Lowering Method}
\begin{figure}[t]
	\begin{center}
		\includegraphics[width=0.48\textwidth]{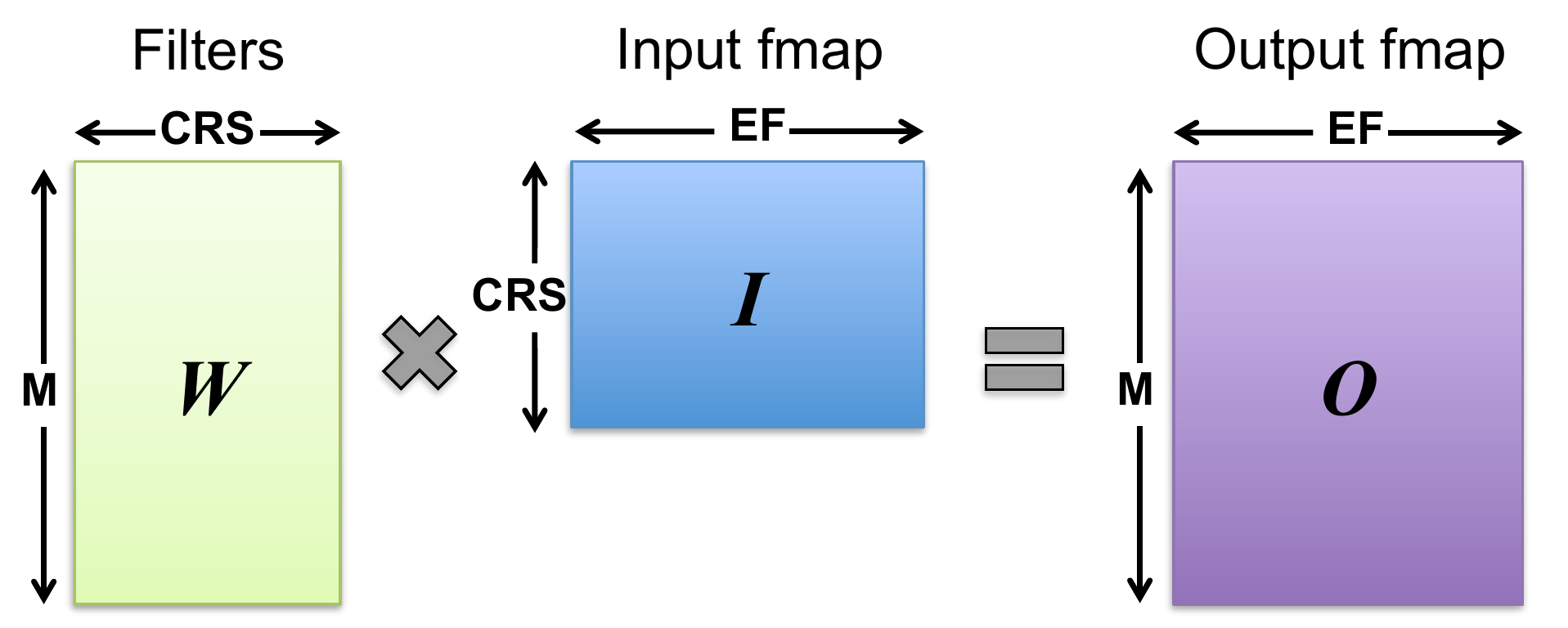}
		\caption{Matrix multiplication is used when computing one 
		output feature map from one input feature map.}
	    \vspace{-0.3cm}
		\label{fig:mm}
	\end{center}
\end{figure}

To leverage highly optimized GEMM (General Matrix Multiply) libraries, CONV layers 
in DNNs are usually mapped to matrix multiplication. \cref{fig:lowering} gives an 
example of transforming 2-D convolution into matrix multiplication. The 2-D filter 
is flattened into a 1-D array, and the input features are filled into a matrix such 
that the dot product of the 1-D array and each column of the matrix generates an 
output element. This process is called the \textit{lowering} method~\cite{cuDNN}.
Extending this process to the 3-D convolution in \cref{fig:cnn}, 
the filters are reshaped into a matrix $W$ with dimensions $M \times CRS$, 
and a input matrix is gathered by duplicating the original input data into 
a matrix $I$ with dimensions $CRS \times EF$. After this transformation, 
the convolution is replaced by a single matrix multiplication in \cref{fig:mm}
to form an output matrix $O$ with dimension $M \times EF$.

There are software libraries designed for GPUs (e.g., cuBLAS) that highly optimize 
matrix multiplications. The implementation is tiled to the memory hierarchy of 
the target GPU to capture locality. Due to the simplicity of implementation and
consistency of performance across the parameter space, the lowering method 
is adopted by most DNN frameworks (e.g. TensorFlow~\cite{TensorFlow}, 
Caffe~\cite{Caffe}, Theano~\cite{Theano}, and Torch7~\cite{Torch7}).

The downside for using GEMM for CONV layers is that there is redundant 
data in the input matrix $I$ as highlighted in \cref{fig:lowering}. This 
can lead to inefficiency in storage and waste of bandwidth at runtime. 
Constructing $I$ requires duplicating the input features up to $R \times S$ 
times, which might require a prohibitively large memory space allocation. 
In this case, implementations (e.g., Caffe) need to materialize $I$ piece 
by piece, e.g., by calling GEMM iteratively for each element of the batch.
However, this approach limits the parallelism, and can lead to cases 
where the matrix multiplications are too small to efficiently utilize the 
GPU~\cite{cuDNN}. Besides, the operation of forming $I$ in memory itself 
is costly, requiring significant memory traffic. More importantly, 
due to duplication, lowering reduces data reuse opportunities and wastes 
memory bandwidth at runtime, which increases the burden of the memory 
subsystem. The lowering approach has demonstrated overhead for dense 
convolution~\cite{CcT}, and unfortunately, this issue gets worse and 
unacceptable when the computation becomes sparse after weight pruning. 

\subsection{Weight Pruning}
Weight pruning techniques~\cite{prune} measure the importance of 
each weight and remove those deemed unimportant, resulting in both 
memory storage and computation reductions with no accuracy loss. 
After weight pruning, redundant weights and related MAC operations 
are removed. One such method, Deep Compression~\cite{DeepCompression} 
can reduce the number of weights in AlexNet~\cite{AlexNet} and 
VGG-16~\cite{VGG-16} by 9$\times$ and 13$\times$, respectively.

\begin{figure}[t]
	\begin{center}
		\includegraphics[width=0.48\textwidth]{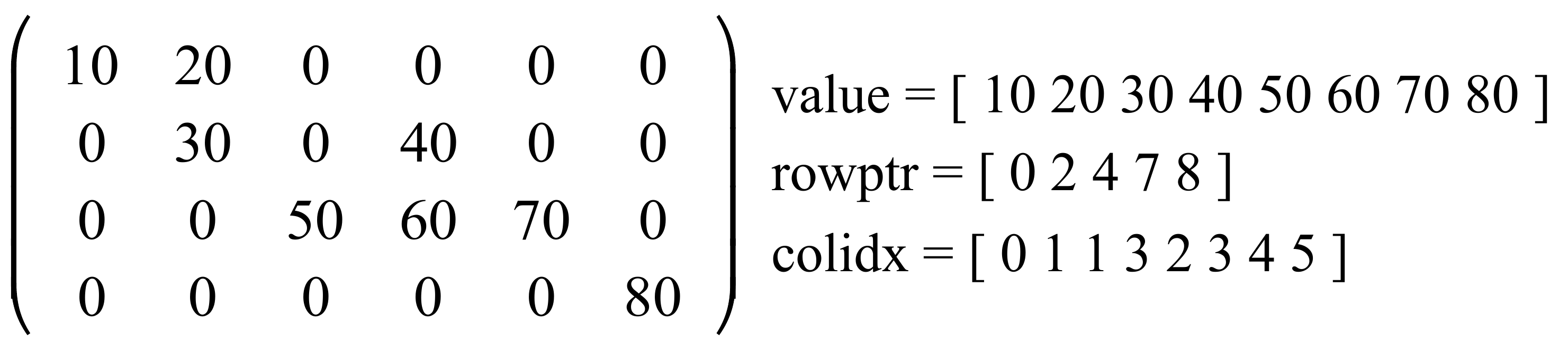}
		\caption{An example of the compressed sparse row (CSR) format.}
		\label{fig:csr}
		\vspace{-0.6cm}
	\end{center}
\end{figure}

After pruning, the remaining weights are stored in a sparse matrix format.
\textit{Compressed sparse row} (CSR) format, as shown in \cref{fig:csr}, 
is often used to store the sparse weight matrix in a compressed form. 
The CSR data structure consists of three arrays. The data array $value$ 
stores only the non-zero elements row by row. To find out the original
location of each non-zero elements, two auxiliary data structures are added.
The column-indices array $colidx$ contains $nnz$ integers ($nnz$ is 
the total number of non-zero elements), and entry $colidx[i]$ indicates 
the column id of the $i$th element in $value$. The row-pointers array $rowptr$ 
contains $M + 1$ ($M$ is the number of rows of the matrix) integers, and 
entry $rowptr[i]$ is the starting index in $colidx$ of the $i$th row. This 
implies that $rowptr[i+1] - rowptr[i]$ is the number of non-zero elements 
in the $i$th row. 

Using CSR format, the memory space used to store the weight matrix is 
$(2\times nnz+M+1)\times 4$ bytes (assuming floating-point data type for 
$value$). We define the \textit{sparsity} of a sparse matrix as the ratio of 
zero values the matrix stores relative to the total number of cells in the matrix.
Since more than 80\% weights are set to zero by the pruning technique, the 
sparsity of the weight matrix is often over 0.8, i.e. $nnz < 0.2 \times total$ 
($total$ is the total number of cells in the matrix), and the memory space 
consumed by the compressed weight matrix is then less than 40\% of the 
original dense matrix (assuming $M \ll nnz$). This can enable deeper CNN 
models in the future, and is also important for deployment of CNN models 
in memory constrained platforms, such as desktops and mobile devices.

\subsection{Limitations of Lowering for Sparse CNN}
Despite the advantage of dramatic reduction of MAC operations, the sparsity 
of pruned networks often leads to performance loss in CNN computation
on GPUs~\cite{Scalpel}. This is because sparse weight matrices lose the 
regular structure of dense matrices. On GPUs, the sparse matrix 
computation~\cite{SpMV,SpMM,SCC} cannot make full usage of the supported 
hardware, e.g., memory coalescing. Also, dense matrix optimizations, 
like matrix tiling, are less effective~\cite{Scalpel}. Therefore sparsity
brings limited benefit if running on GPUs. Worse still, extra overhead
is needed to decode the sparse format at runtime. With limited benefit, it
is not surprising that this overhead would lead to performance degradation.

\cref{fig:speedup-sconv} illustrates execution time spent on CONV layers when 
performing inference on NVIDIA GPUs using CUBLAS and CUSPARSE respectively. 
For both methods, the weight matrices are pruned, but they are stored as dense 
matrices (filled with lots of zeros) for CUBLAS and as sparse matrices (i.e. CSR) 
for CUSPARSE. We can observe a consistent performance  loss on the Tesla P100 GPU. 
As for GTX 1080Ti GPU, CUSPARSE achieves very limited performance improvement 
compared to CUBLAS. This unsatisfactory performance motivates us to rethink the 
mapping of convolution operations to GPUs and optimize the implementation for 
the data-parallel architecture.

\subsection{GPU Programming and Memory Hierarchy}
From the programmers' point of view, each CUDA kernel includes groups of threads 
called {\it thread blocks}. All threads in a thread block are guaranteed to execute 
concurrently on the same streaming multiprocessor (SM). Within each thread block, 
subgroups of threads called {\it warps} (usually containing 32 threads) are executed in 
{\it lockstep} fashion. This programming paradigm is defined to fit GPU's SIMT 
architecture~\cite{CUDA}. When a multiprocessor is given one or more thread blocks 
to execute, it partitions them into warps and each warp gets scheduled for execution 
on the SIMD execution units.

The GPU memory hierarchy 
consists of several levels of storage with variable sizes, properties, and access constraints. 
Register files are the closest to the streaming multiprocessor, and they are local memories
for each thread. Shared memory, a.k.a. scratchpad, is programmer manageable and can be 
shared by the threads in the same thread block. At the same level there is a hardware-managed
read-only cache, which is used to hold the read-only data specified by the programmer. 
The L2 cache is shared across all threads of the entire CUDA kernel and usually works as 
the central point of coherency. Besides, memory requests would reach off-chip GDDR or 
HBM2 DRAM when the required data is not in any of the above levels~\cite{P100}. 

\section{Escort Design}\label{sect:design}
As mentioned, the lowering approach replicates the input features multiple times, 
significantly reducing arithmetic intensity, and this issue is particularly worse for
sparse convolution since its intensity is already much lower than that of dense one. 
To avoid this limitation, we use the \textit{direct sparse convolution} 
method~\cite{SkimCaffe} to perform convolution. We then map the operations onto 
GPUs with SIMT parallelism in mind. We also analyze the memory access pattern of 
sparse convolution and employ optimization techniques to improve data locality. 
For various layers with different parameters (e.g. the sizes of filters and ofmaps), 
we adaptively apply customized compute kernels to improve efficiency. 

\begin{algorithm}[!t]
	\caption{Sequential Sparse Convolution~\cite{SkimCaffe}}
	\label{alg:sconv}
	\begin{algorithmic}[1]
		\Procedure{SConv}{in, W, out}
		\For{$n$ in [0, N)}
		\For{$m$ in [0, M)}
		\For{$j$ in [W.rowptr[$m$], W.rowptr[$m$+1])}
		\State \texttt{off} $\leftarrow$ W.colidx[$j$]
		\State \texttt{val} $\leftarrow$ W.value[$j$]
		\For{$h$ in [0, E)}
		\For{$w$ in [0, F)}
		\State out[$n$][$m$][$h$][$w$] += \texttt{val} * \\
		\hspace{3.5cm} in[$n$][\texttt{off} + $f(0,h,w)$]
		\EndFor
		\EndFor
		\EndFor
		\EndFor
		\EndFor
		\EndProcedure
	\end{algorithmic}
\end{algorithm}

\subsection{Overview}
For the lowering method, materializing the lowered matrix in memory can 
be costly for GPUs whose memory size is relatively limited. To avoid this 
overhead, cuDNN materializes the lowered matrix by lazily loading the 
input matrix into on-chip cache at runtime, rather than by constructing 
it in off-chip memory before calling a GEMM routine~\cite{cuDNN}. 
Escort follows this approach, but adapts it for direct sparse convolution. 
A 1-D array is used to hold the ifmaps, padded if necessary. 
As the computation proceeds, we dynamically compute the offset 
of the input array, and then use the index to load the correct 
elements into on-chip memories. After the computation is complete, 
we perform the required index calculation to store the result in the correct 
output position. We refer this technique as \textit{dynamic indexing}. 

In SkimCaffe~\cite{SkimCaffe}, a layout function $f$ is defined such 
that $f(c,y,x)$ maps to the offset corresponding to $(c,y,x)$th element 
of the input array (assuming that $f(c,y+r,x+s)=f(c,y,x)+f(0,r,s)$). 
For example, in CHW layout, $f (c,r,s) = (c \cdot H_{in}+r)W_{in}+s$. 
We use this function $f$ to compute the correct index of the input array.
The \textit{dynamic indexing} approach improves arithmetic intensity, 
at the cost of dynamically calculating the index of input array. This 
trade-off is made based on the fact that sparse computation is often 
highly memory bound and it is important to reduce off-chip memory accesses 
to improve GPU efficiency, even at the cost of more index calculation.

To match the dimension of the input array, the weight matrix is
stretched beforehand. This is preprocessed when constructing the 
sparse weight matrix (i.e. the CSR data structures) and only run once. 
We refer this preprocessing operation as \textit{weight stretching}~\cite{SkimCaffe}.
This operation only modifies the column indices of the weight matrix
which are stored in the $colidx$ array. No extra memory space is consumed.

\begin{figure}[b]
	\begin{center}
		\includegraphics[width=0.48\textwidth]{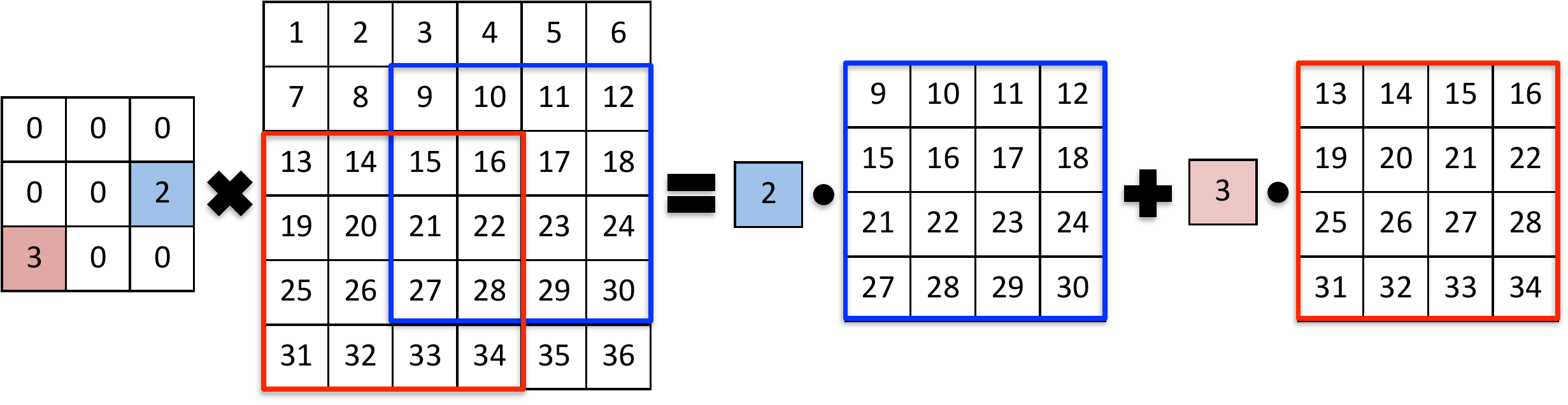}
		\caption{An example sparse convolution of one 3$\times$3 filter 
			against an 6$\times$6 input feature with 1 channel~\cite{SparseCNN}.}
		\label{fig:dataflow}
		\vspace{-0.33cm}
	\end{center}
\end{figure}

The sequential algorithm of direct sparse convolution is calculated as shown 
in Algorithm~\ref{alg:sconv}. For each ofmap (line 2) and each output channel 
in the ofmap (line 3), it traverses all the elements in the corresponding filter 
(line 4), and gets the offset (line 5) and weight value (line 6) from the CSR data
structures. It then iterates over the 2-D channel in row-major order (line 7\&8). 
At last it loads the input feature using the dynamically calculated index, multiplies 
it with the weight value, and accumulates the product to the correct output location (line 9\&10).

\begin{figure}[t]
	\begin{center}
		\includegraphics[width=0.33\textwidth]{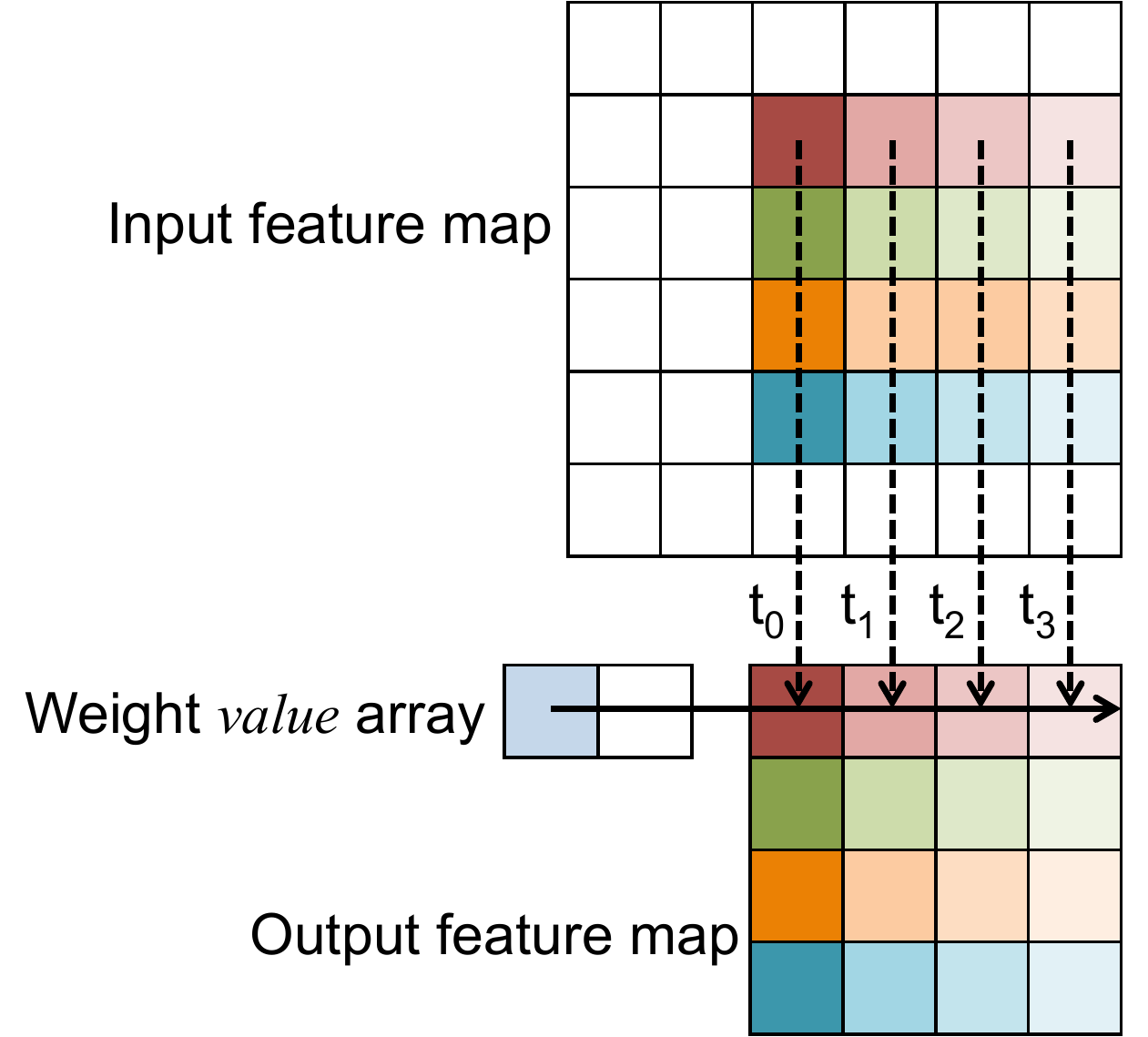}
		\caption{Data-to-thread mapping for a pseudo 4-thread warp.}
		\label{fig:mapping}
		\vspace{-0.5cm}
	\end{center}
\end{figure}

\subsection{Parallelism Strategy}
A CNN's dataflow defines how the loops are ordered, partitioned, and parallelized~\cite{Eyeriss}.
A straightforward implementation of Algorithm~\ref{alg:sconv} is not necessarily 
efficient on GPUs if the dataflow is not carefully designed for the underlying architecture.
For example, non-contiguous indirect memory access is a major overhead of typical 
sparse-matrix computations on GPUs~\cite{SpMV}. If consecutive threads in a warp 
accesses consecutive memory locations, the memory requests are coalesced into one 
or several memory transactions to save memory bandwidth. Otherwise, memory divergence 
occurs and the efficiency of GPU memory subsystem declines sharply~\cite{Chen}.

We choose a dataflow in~\cite{SparseCNN} to minimize memory divergence. The basic 
idea is illustrated in \cref{fig:dataflow}. The sparse convolution of a 3$\times$3 
filter against a 6$\times$6 input feature can be divided into two parts: the nonzero 
weight ``2'' in the filter times a 4$\times$4 sub-matrix (blue), and the other 
nonzero weight ``3'' in the filter times another 4$\times$4 sub-matrix (red). 
And then the final results is obtained by simply accumulating the two products. 
Unstructured computation is avoided when separately conducting the multiplications.  

The data-to-thread mapping on GPU is shown in \cref{fig:mapping}. Assuming a 4-thread 
warp, the accesses to the input array by a warp are coalesced as long as the 
array elements with contiguous row or column indices are stored contiguously. 
Each thread is responsible for calculating one output element in the output matrix. 
When writing the product sum into the output array, the accesses are also contiguous 
since consecutive threads are assigned to calculate consecutive output positions. 
Thus for each non-zero weight, it is multiplied with consecutive input data in the 
same row, and each product is added to the partial sum of the corresponding output 
element which is assigned to the thread. In this way, we can avoid most of uncoalesced 
memory accesses to the global memory in the GPU, and thus improve memory access efficiency.

\subsection{Locality}
The key to highly efficient sparse convolution on GPUs is to maximize data locality.
Previous research has shown that the overall performance of memory intensive 
applications on GPU is highly affected by its on-chip cache performance~\cite{Chen}.
To have a deep understanding of the reuse pattern of sparse convolution, we analyze 
the nested loop in Algorithm~\ref{alg:sconv}. It can be transformed in numerous ways 
to capture different reuse patterns of the weights and activations, and to map the 
computation to the underlying hardware~\cite{SCNN}. For example, an input channel is 
reused against multiple filters to generate multiple output channels, and there is 
also ample reuse out of an input channel due to overlapping between sliding windows. 
A filter is reused not only when it is sliding across an ifmap, but also against 
multiple ifmaps in a batch. Thus the arithmetic intensity of sparse convolution is 
significantly higher than typical sparse-matrix computations. Also, potential data 
reuses in direct sparse convolution are more than that in lowered GEMM, since some 
reuses are lost when duplicating the input features. This implies that it is possible 
to achieve high compute efficiency on GPUs.

\begin{figure}[t]
	\begin{center}
		\includegraphics[width=0.46\textwidth]{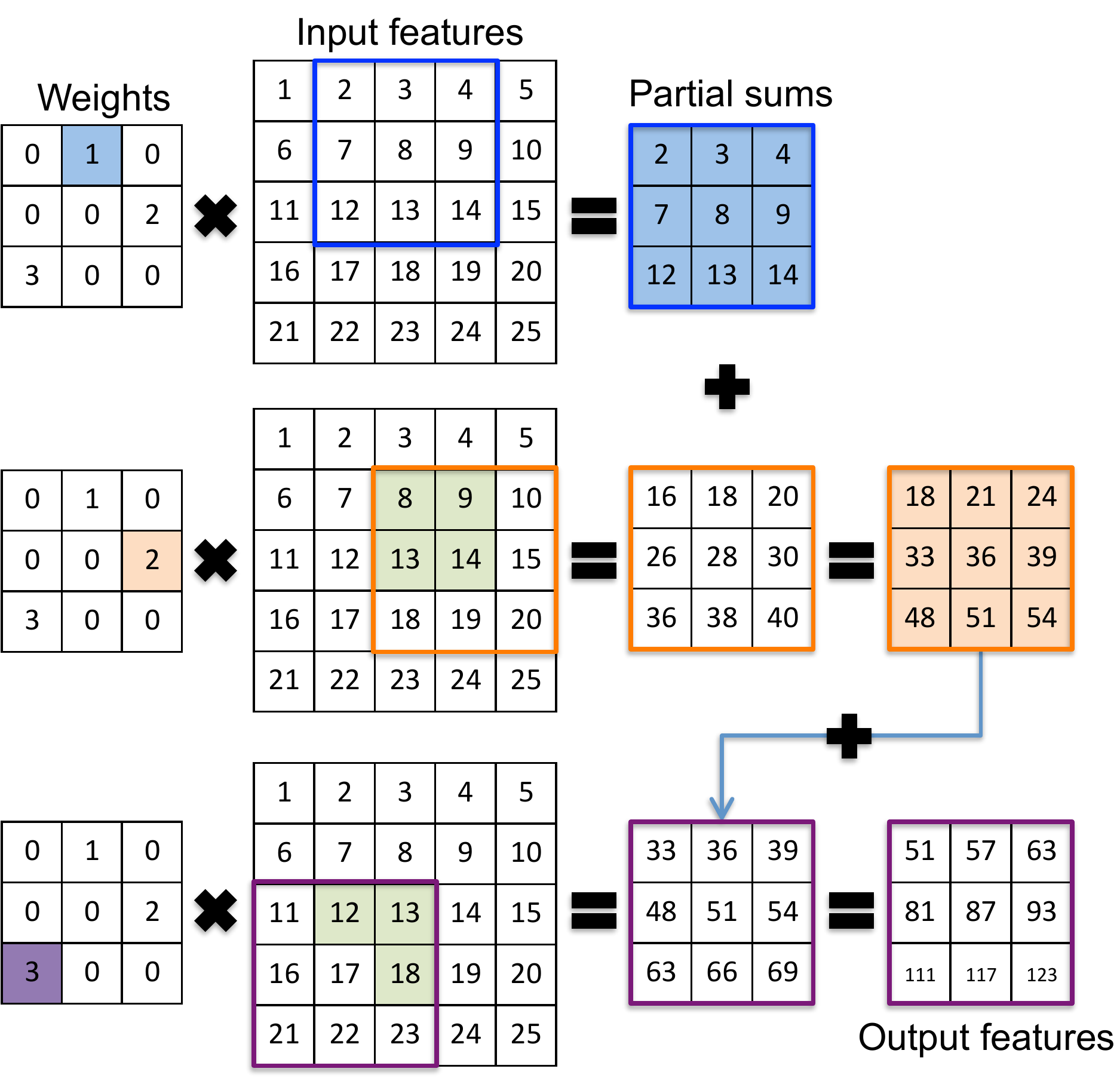}
		\caption{An example of data reuse in sparse convolution. 
			The colored boxes represents the data that is accessed multiple times.}
		\label{fig:reuse}
		\vspace{-0.6cm}
	\end{center}
\end{figure}

In Algorithm~\ref{alg:sconv}, three major data structures are used in the sparse convolution: 
the input features, the sparse weight matrix (CSR format), and the output features. 
Therefore, generally we have three types of dataflow to capture reuse~\cite{Tutorial}: 
1) \textit{Weight Stationary} is to minimize the overhead of loading weights by maximizing 
the accesses of weights in the on-chip cache. 2) \textit{Output Stationary} aims to 
minimize the overhead of reading and writing the partial sums. It keeps the accumulation 
of partial sums in the on-chip cache, and streams the input features across the processor 
and broadcasts the weights, and 3) \textit{Input Stationary} is to minimize the overhead 
of reading inputs by keeping the input features in the cache and streams the weights.

We try to maximize the reuse and accumulation in the cache for all types of data, i.e., weights, 
inputs and partial sums. We assign the work of processing one output channel to a thread block. 
It keeps the corresponding filter weights stationary inside the cache, and then streams 
the input features into the SM. Since there are overlaps of input features between 
different sliding windows, the input features are also be kept in the cache and get reused. 
\cref{fig:reuse} shows an example of the data reuse captured when calculating sparse convolution. 
In this case, each element read from the sparse weight matrix is reused $E \times F$ times.

To fully exploit data locality, we should carefully arrange the data placement, i.e. 
find the suitable kinds of memory to place different types of data. The input features 
and the weight matrix are read-only, while the output features are written. Since the 
weight matrix is stored as CSR format, we use threads in a thread block to cooperatively 
load the $colidx$ and $value$ arrays into the shared memory. These are all coalesced 
memory accesses. Since the input data is not modified throughout the entire process, 
we put them in the read-only cache so that they can be shared across thread blocks 
running on the same SM and reused multiple times. As for the partial sums, they are 
put in the register files to keep the accumulation local and fast.

\subsection{Kernel Customization}
Implementations following the direct sparse convolution approach should be specifically 
optimized for convolutions in certain parts of the parameter space. The major factors
we should consider includes the filter size, the ofmap size, the batch size and the stride.
We use C++ template to generalize the kernel source, and let the compiler dose the work
of generating customized kernel for specific parameters. The optimization space we explore
includes the grid shape and thread block size. Besides, to improve performance when the 
filter size is smaller than 3$\times$3, cuDNN uses Winograd~\cite{Winograd} algorithm 
instead of lowering onto matrix multiplication to perform convolution. 
This approach is compatible with Escort. We take this as a future work.
\section{Evaluation}\label{sect:eval}

We evaluate performance of Escort on two platforms shown in Table~\ref{table:platform}. 
NVIDIA Tesla P100~\cite{P100} represents data-center server platform. NVIDIA Geforce 
GTX 1080Ti represents desktop environment. Escort is implemented as an extension 
of Caffe deep learning framework~\cite{Caffe}. We use gcc 4.8 and NVCC 8.0 for compilation. 
\texttt{nvprof}~\cite{NVVP} is used to collect execution time and performance metrics of 
CUDA kernels. All the experiments use 32-bit floating point data type and batch size of 128. 
We use trained and pruned models of AlexNet~\cite{AlexNet}, GooLeNet~\cite{GooLeNet}, and 
ResNet~\cite{ResNet} which are available in the SkimCaffe repository~\cite{repo} (along with the 
sparsity information). All these models are trained on the ImageNet~\cite{ImageNet} ILSVRC-2012 
dataset. Details about the models are listed in Table~\ref{table:nets}. Since optimizations 
used in Escort has no effect on accuracy, our evaluation focuses on performance (i.e. inference speed).
\renewcommand\theadalign{bc}
\renewcommand\theadfont{\bfseries}
\renewcommand\theadgape{\Gape[4pt]}
\renewcommand\cellgape{\Gape[4pt]}

\begin{table}[b]
	\footnotesize
	\centering
	\begin{tabular}{c c c}
		\hline
		\hline
		& \texttt{GTX 1080Ti} & \texttt{Tesla P100}\\
		\hline
		\textbf{\# of cores} & 3584 & 3584\\
		\hline
		\textbf{Boost Clock} & 1582 MHz & 1480 MHz\\
		\hline
		\textbf{Mem. Size} & 11 GB GDDR5X & 16 GB HBM2\\
		\hline
		\textbf{Bandwidth} & 484 GB/s & 732 GB/s\\
		\hline
		\hline
	\end{tabular}
	\caption{Evaluated GPU Platforms}
	\label{table:platform}
\end{table}

\begin{table}[b]
	\footnotesize
	\centering
	\begin{tabular}{c c c c c}
		\hline
		\hline
		\thead{Model} & \thead{\# of CONV \\Layers} & \thead{\# of Sparse\\ CONV Layers} & \thead{Weights} & \thead{MACs}\\
		\hline
		\texttt{AlexNet} & 5 & 4 & 61M & 724M\\
		\hline
		\texttt{GoogLeNet} & 57 & 19 & 7M & 1.43G\\
		\hline
		\texttt{ResNet} & 53 & 16 & 25.5M & 3.9G\\
		\hline
		\hline
	\end{tabular}
	\caption{Summary of Networks}
	\vspace{-0.3cm}
	\label{table:nets}
\end{table}

\begin{figure}[t]
\begin{center}
	\includegraphics[width=0.48\textwidth]{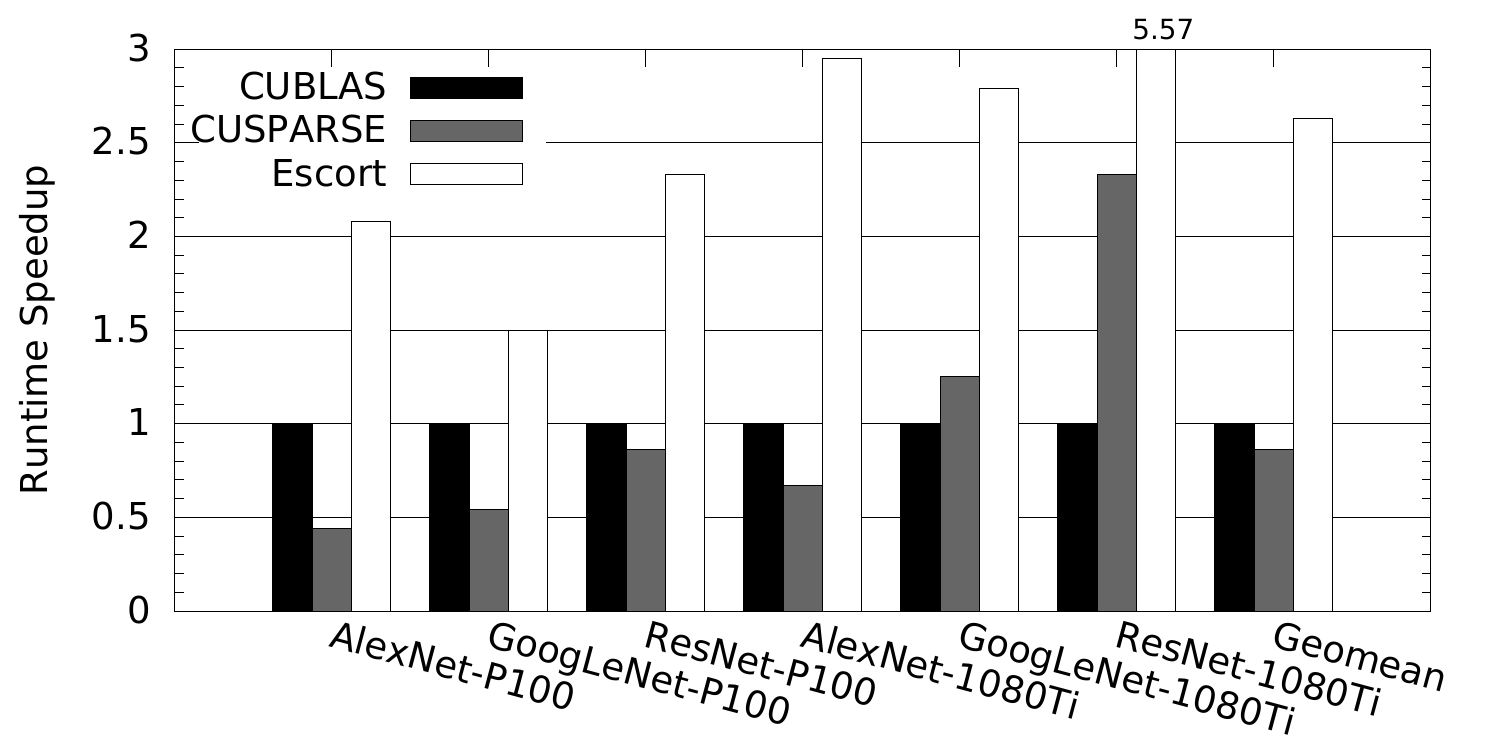}
	\caption{Execution time speedup of sparse CONV layers in three 
		models all normalized to the CUBLAS approach. Dense CONV layers 
		and other layers are not included in this experiment.}
	\label{fig:speedup-sconv}
	\vspace{-0.5cm}
\end{center}
\end{figure}

\subsection{Sparse CONV Performance}
Firstly, we compare the performance of sparse CONV layers in CUBLAS, 
CUSPARSE and Escort. \cref{fig:speedup-sconv} shows the normalized 
execution time of the sparse CONV layers in these three implementations. 
We collect the timing information using \texttt{nvprof}. We only accumulate 
execution time related to sparse CONV layers, i.e., the execution time 
spent on dense CONV layers and other non-CONV layers (such as FC, ReLU, LRN 
and Pooling layers) is not collected. We can observe that CUSPARSE on Tesla 
P100 suffers a consistent performance degradation compared to CUBLAS due to 
the irregularity of the sparse kernels in CUSPARSE, while on GTX 1080Ti, 
CUSPARSE can accelerate GoogLeNet and ResNet by 1.25$\times$ and 2.33$\times$, 
but still causes slowdown for AlexNet. On the contrary, Escort consistently 
achieves significant performance improvement over CUBLAS, with speedups from 
1.50$\times$ and 5.57$\times$. On average, sparse CONV layers in Escort is 
2.63$\times$ and 3.07$\times$ faster than those in CUBLAS and CUSPARSE respectively.

\subsection{Execution Time Breakdown}

\begin{figure}[t]
	\begin{center}
		\includegraphics[width=0.48\textwidth]{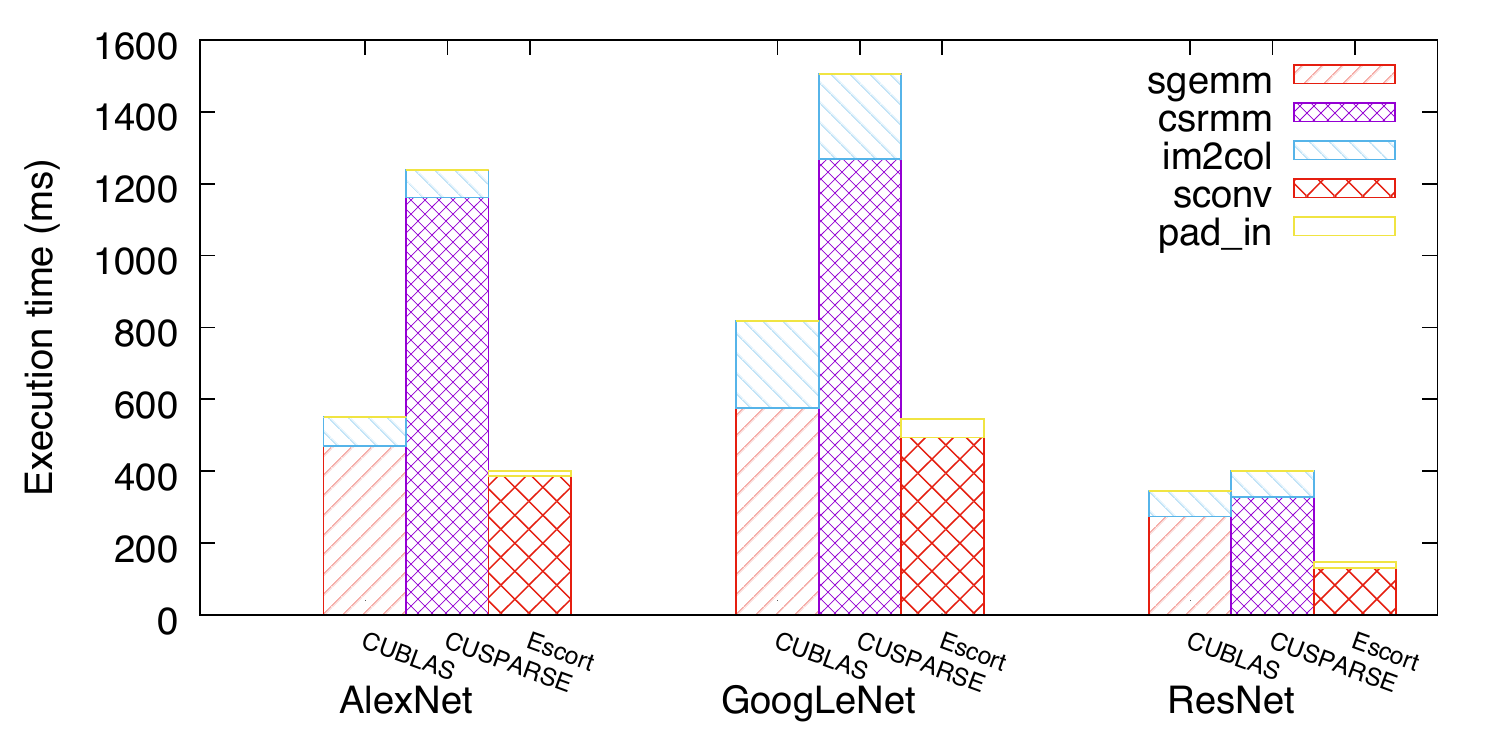}
		\caption{Execution Time Breakdown for Sparse CONV Layers.}
		\label{fig:breakdown}
		\vspace{-0.5cm}
	\end{center}
\end{figure}

To investigate the performance effect in detail, we breakdown the execution time of
sparse CONV layers on Tesla P100 into several parts, each of which is a CUDA kernel.
The kernel timing is collected using \texttt{nvprof}. The kernels include: \texttt{sgemm}, 
\texttt{csrmm}, \texttt{im2col}, \texttt{sconv} and \texttt{pad\_in}. \texttt{sgemm} 
is the dense matrix multiplication routine in CUBLAS. \texttt{csrmm} is the sparse 
matrix dense matrix multiplication routine in CUSPARSE. \texttt{im2col} 
is the CUDA kernel to lower the input data onto matrices. 
\texttt{sconv} is CUDA kernel of our proposed sparse convolution in Escort. 
\texttt{pad\_in} is the kernel that Escort uses to pad the input data.
\cref{fig:breakdown} shows the execution time distribution of different CNN models 
using different approaches. Since CUBLAS and CUSPARSE are both base on the lowering
method, they have the same execution time spent on \texttt{im2col}. Escort does not
require this data transformation, and the input padding process \texttt{pad\_in} is 
less costly than \texttt{im2col}. As for the core computation part, \texttt{sgemm}
is faster than \texttt{csrmm}, due to the irregularity of sparse matrix computation.
However, \texttt{sconv} is faster than \texttt{sgemm}, which demonstrates the
effectiveness of our optimization techniques.



\begin{figure}[t]
	\begin{center}
		\includegraphics[width=0.48\textwidth]{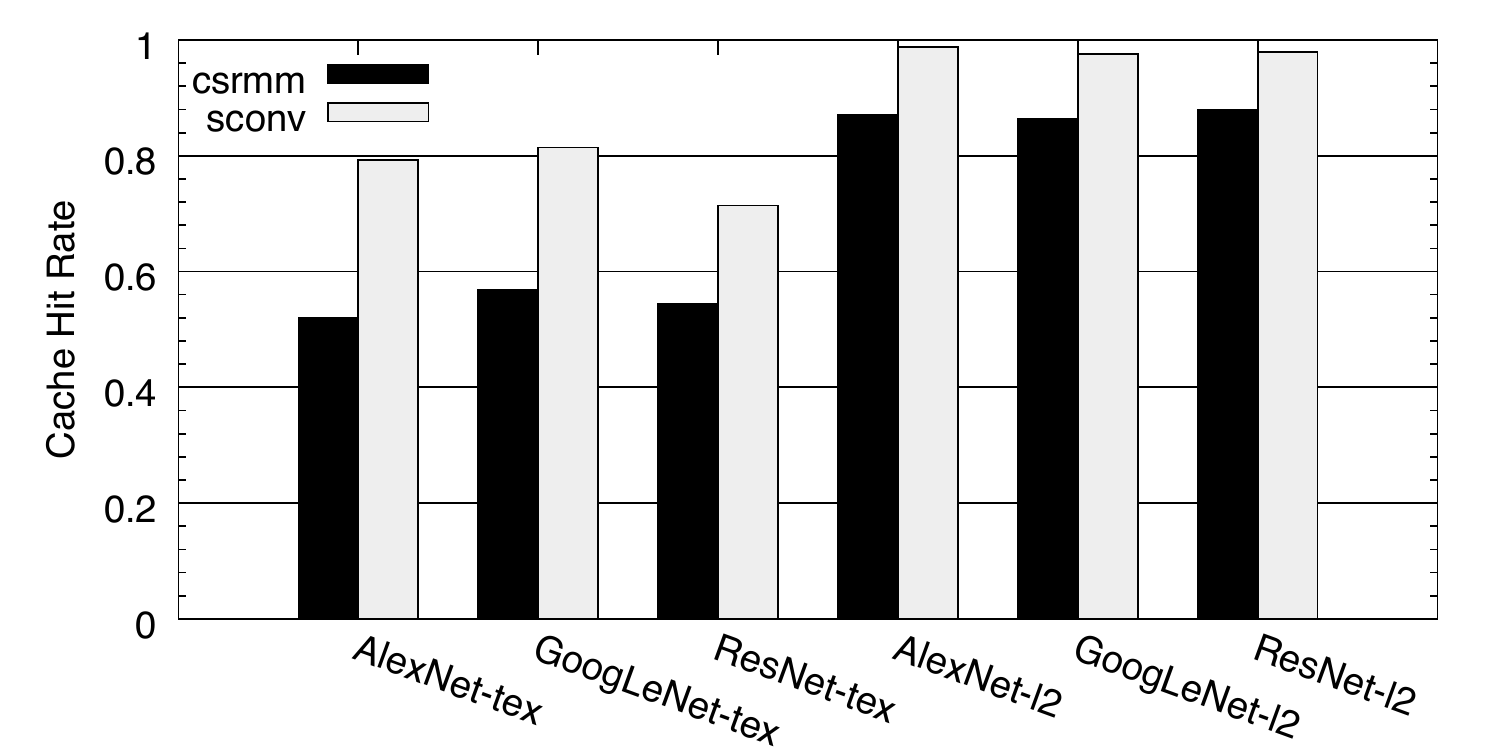}
		\caption{Cache hit rate of two CUDA kernels on Tesla P100.}
		\label{fig:cache-hit-rate}
		\vspace{-0.6cm}
	\end{center}
\end{figure}

\subsection{On-chip Memory Efficiency}
\cref{fig:cache-hit-rate} compares the texture (i.e. read-only) cache and L2 cache hit rate of two 
CUDA kernels, \texttt{csrmm} and \texttt{sconv}. The results are collected by \texttt{nvprof}.
For all three models, \texttt{sconv} in Escort consistently achieves better read-only cache 
performance (hit rates from 71\% to 81\%) compared to \texttt{csrmm} in CUSPARSE 
(hit rates from 52\% to 57\%). As for L2 cache, we observe similar trend. This is reasonable 
because cache tiling is not as effective for sparse matrix computation as its dense counter 
part~\cite{Scalpel}, and some data reuses have already been lost when duplicating the input 
features. In contrast, we separately store the weight and input features in different kinds of 
on-chip memories, avoiding possible cache conflicts, and adaptively tile the output channel 
to make good use of the read-only cache.

\subsection{Overall Performance}
\begin{figure}[t]
\begin{center}
	\includegraphics[width=0.48\textwidth]{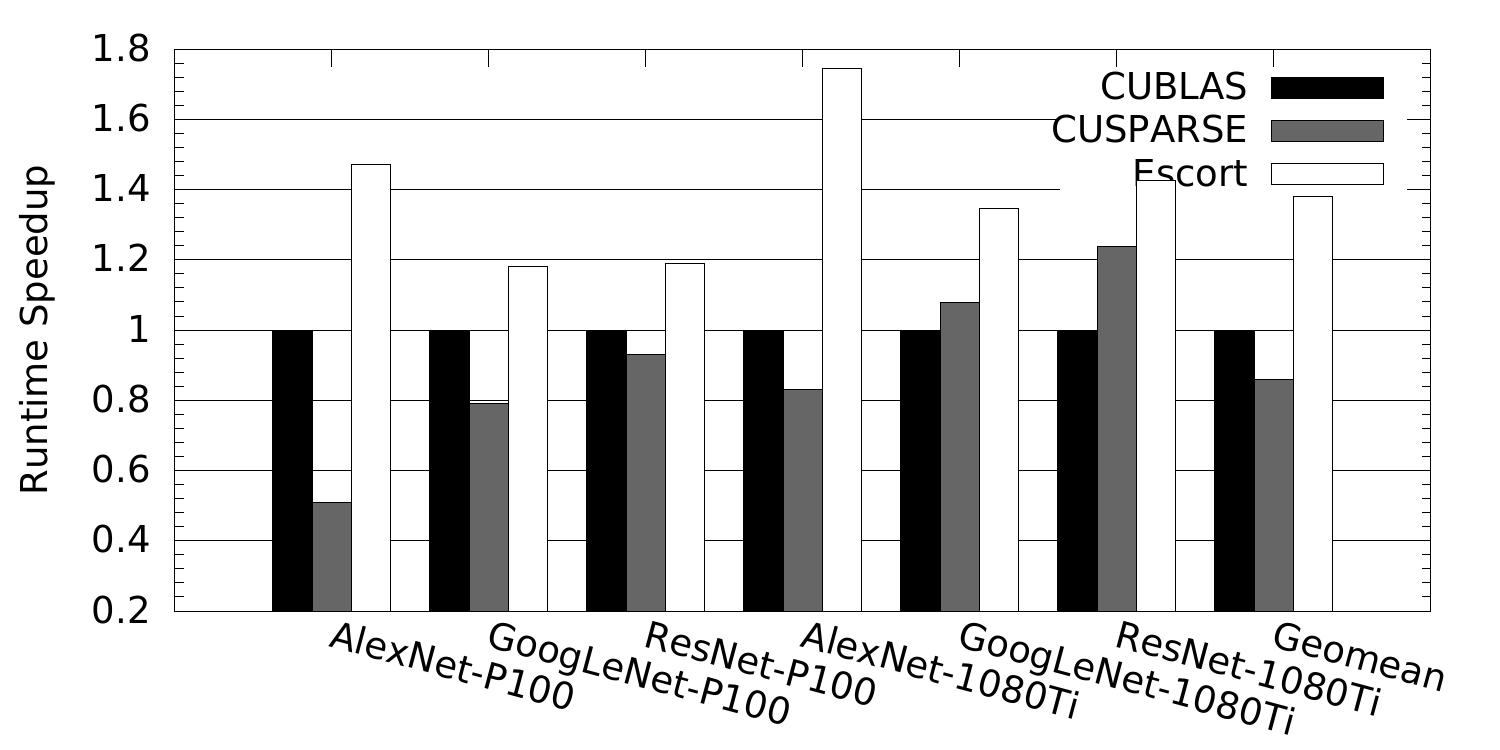}
	\caption{Overall performance speedup of three models,
		all normalized to the CUBLAS approach.}
	\label{fig:speedup}
	\vspace{-0.6cm}
\end{center}
\end{figure}

\cref{fig:speedup} illustrates the overall inference performance of the three approaches.
In this experiment, we collect the execution time spent on an entire iteration (i.e the 
time spent on processing one batch) in Caffe. To avoid noise, we run 10 iterations and
calculate the average time. We observe similar trend as \cref{fig:speedup-sconv}, but the 
performance variation among different approaches is less significant since we add 
up the execution time of all the other layers. Even so, Escort still achieves consistent 
speedup over CUBLAS, i.e., 1.47$\times$, 1.18$\times$ and 1.19$\times$ on Tesla P100, 
and 1.74$\times$, 1.34$\times$ and 1.43$\times$ on GTX 1080Ti, for AlexNet, 
GoogLeNet and ResNet respectively. Escort gets the smallest speedup for GoogLeNet 
because a large portion of CONV layers are dense and can not benefit from our sparse
convolution method. As for ResNet, the performance of CUSPARSE and Escort is not as 
significantly affected as that of AlexNet because of its relatively lower proportion of CONV 
layers in all layers. On average, Escort achieves a geomean speedup of 1.38$\times$
over the CUBLAS approach which is the default GPU configuration of Caffe. Compared to 
CUSPARSE, the speedup is 1.60$\times$. Note that this performance improvement requires 
neither adaption of higher level programming nor modification of underlying hardware.

\section{Related Work}\label{sect:relate}

\textbf{Sparse CNN on CPUs}.
Liu~\emph{et~al.}~\cite{Liu} proposed a sparse dense MM algorithm for inference 
on CPUs, which exploits sparsity in the weights. Park~\emph{et~al.}~\cite{SparseCNN} 
implemented direct sparse convolution for inference on CPUs and optimizes it for
different kinds of Intel CPUs. Meanwhile, Rajbhandari~\emph{et~al.}~\cite{OCM} 
proposed to leverage sparsity for training DNNs on CPUs and develops an optimization 
framework to automatically choose best performing implementations for various CNN 
computations. Vooturi~\emph{et~al.}~\cite{Vooturi} proposed parallel algorithms to 
perform efficient inferencing on multicore CPUs using MKL. These experiences provide 
us insights for implementing sparse CNN on GPUs.

\textbf{Sparse CNN Accelerators}.
Recent works have examined how to efficiently support processing of sparse CNN in hardware. 
EIE~\cite{EIE} compresses the model in the fully connected layers to speedup inference. 
Eyeriss~\cite{Eyeriss} gates the multiplier when the input activation is zero, while
Cnvlutin~\cite{Cnvlutin} compresses activation values to skip over the ineffectual computations. 
But neither of them leverage pruning to exploit weight sparsity. Cambricon-X~\cite{Cambricon-X} 
employs weight sparsity to keep only non-zero weights in its internal buffers. SCNN~\cite{SCNN} 
keeps both weights and activations in a compressed form and uses Cartesian product to compute 
convolution. Comparing with these hardware solutions, Escort is a pure software approach and
requires no effort from either high level programmers or hardware designers.

\textbf{Structured Pruning}.
Recent works have explored the use of structured pruning to regularize sparse 
matrix computation on GPUs. Structured Sparsity Learning (SSL)~\cite{SSL} 
adaptively regularizes DNN structures, and employs locality optimization 
to accelerate computation. Scalpel~\cite{Scalpel} leverages SIMD-aware 
weight pruning and node pruning for CPUs and GPUs respectively. 
DeftNN~\cite{DeftNN} presents synapse vector elimination and applies a 
transformation to the DNN data layout, producing efficient computations on GPUs. 
Molchanov~\emph{et~al.}~\cite{Molchanov} proposed to prune filters to 
enable efficient inference. Mao~\emph{et~al.}~\cite{Regularity} compared 
different kinds of pruning techniques at different pruning granularities.
Compared with these structured pruning approaches, Escort directly improves
performance on arbitrary sparse networks, requiring no adjustment of the
training and pruning process, and it has no effect on the inference accuracy.


\section{Conclusion}\label{sect:concl}
CNNs have been applied in a wide range of AI applications and achieved 
remarkable performance. To enable deeper and more complex neural networks 
on various platforms, e.g. mobile devices, weight pruning is proposed to 
remove redundant parameters. Unfortunately, pruning 
generates unstructured sparse matrices and leads to unsatisfactory inference 
speed on GPUs which are suited for accelerating structured compute kernels. 
To handle the irregularity of sparse computation, we propose Escort, an 
efficient sparse convolution method customized for GPUs. Escort improves
arithmetic intensity by directly computing sparse convolution instead of 
lowering it onto matrix multiplication, and is specifically optimized for the 
GPU architecture by orchestrating the parallelism and exploiting data locality. 
Our evaluation demonstrates that Escort outperforms the lowering method
implemented on top of either CUBLAS or CUSPARSE, successfully turning 
sparsity into inference speedup on GPUs, not only memory space saving.

\bibliographystyle{acmart}
\bibliography{references}
\end{document}